\documentclass[aps,twocolumn,keywords]{revtex4-1}
\usepackage{amsmath,amsthm,amssymb,graphicx,xcolor}
\usepackage[unicode=true]{hyperref}
\usepackage{amsmath}
\usepackage{rotating}
\usepackage[ruled]{algorithm2e}
\usepackage{extarrows}
\usepackage{graphicx}
\usepackage{dcolumn}
\usepackage{bm}
\usepackage[all]{xy}
\usepackage{indentfirst}
\usepackage{amsmath}
\usepackage{bbm}
\usepackage{multirow}
\usepackage{mathrsfs}
\usepackage{xcolor}
\usepackage{euscript}
\usepackage{amssymb}

\begin{document}
\title{Genuinely Multipartite Entanglement vias Shallow Quantum Circuits}

\author{Ming-Xing Luo}
\email{mxluo@swjtu.edu.cn}
\affiliation{School of Information Science and Technology, Southwest Jiaotong University, Chengdu 610031, P.R. China}

\author{Shao-Ming Fei}
\affiliation{School of Mathematical Sciences, Capital Normal University, Beijing 100048, P.R. China}
\affiliation{Max-Planck-Institute for Mathematics in the Sciences, 04103 Leipzig, Germany}
	
\begin{abstract}
Multipartite entanglement is of important resources for quantum communication and quantum computation. Our goal in this paper is to characterize general multipartite entangled states according to shallow quantum circuits. We firstly prove any genuinely multipartite entanglement on finite-dimensional spaces can be generated by using 2-layer shallow quantum circuit consisting of two biseparable quantum channels, which the smallest nontrivial circuit depth in the shallow quantum circuit model. We further propose a semi-device-independent entanglement model depending on the local connection ability in the second layer of quantum circuits. This implies a complete hierarchy of distinguishing genuinely multipartite entangled states. It shows a completely different multipartite nonlocality from the quantum network entanglement. These results show new insights for the multipartite entanglement, quantum network, and measurement-based quantum computation.
\end{abstract}


\maketitle

\section*{Introduction}

Entanglement is an important quantum property of two or more systems in quantum mechanics associated with Schr{\"o}dinger evolution equations \cite{Schr}. A bipartite entanglement is defined as its cannot be decomposed into an ensemble of separable states. This allows verifying any bipartite entanglement beyond all separable states using entanglement witness \cite{HHH}. Another device-independent method is inspired by Einstein-Podolsky-Rosen (EPR) steering \cite{EPR} or Bell inequality \cite{Bell,CHSH,Brun,Luo1,TPL} which can witness stronger quantum nonlocalities from only the statistics of local measurements on an entanglement \cite{Wise}.

Although the fully separable model is easily extended for multi-particle systems, it is useless for verifying  the genuinely multipartite nonlocality. Instead, some stronger separable models are constructed for special goals. One is the so-called biseparable model \cite{Sv} which is defined to distinguish the genuinely multipartite entanglement from all the biseparable states. This allows us to witness  the genuine multipartite nonlocality beyond the fully separable model even for quantum networks \cite{Luo22}. Another model is using GHZ-paradox \cite{GHZ,tang,Liu,Luois} in the all-versus-nothing test manner. If the local tensor decomposition is considered, the high-dimensional model \cite{Kraft18,Des,EKZ} or quantum network entanglement may rule out any network separable state consisting of small entanglements that are shared by partial parties \cite{NWR,Kraft,Luo2021}. Different from the biseparable model \cite{Sv}, this model provides a device-independent verification of unknown entanglement devices. Another is from the particle-losing model for characterizing the entanglement robustness against losing partial systems \cite{QABZ,BPB,Luo2022}. This can imply a different hierarchy of well-known entanglements including GHZ state and Dicke states going beyond other models \cite{Sv,NWR,Kraft,Luo2021}. All of these entanglement models can only justify special multipartite systems. A natural problem is to explore proper model for general systems.

\begin{figure}[ht]
\begin{center}
\resizebox{200pt}{160pt}{\includegraphics{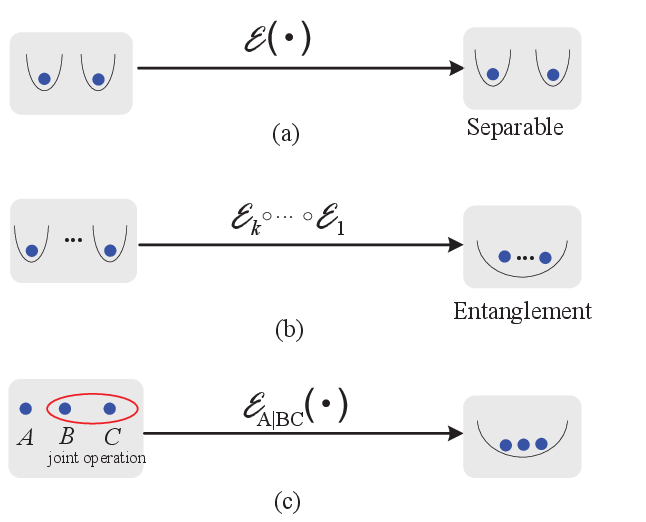}}
\end{center}
\caption{Schematic shallow quantum circuit for quantum state generations. (a) A bipartite separable system with a 1-layer quantum circuit. $\mathcal{E}(\cdot)$ is a bipartite separable channel defined by $\mathcal{E}(\cdot)=\sum_iK_i\otimes S_i\left(\cdot\right) K_i^\dag\otimes S_i^\dag$, where Kraus operators $K_i$ and $S_i$ satisfy $\sum_iK_i^\dag{}K_i\otimes S_i^\dag S_i=\openone$ with the identity operator $\openone$. (b) An $n$-partite entanglement with $k$-layer quantum circuits. Each $\mathcal{E}_j(\cdot)$ is a biseparable quantum channel. (c) A tripartite entanglement with 2-layer quantum circuits. One is from a biseparable quantum channel and the other is from a local joint model in adversarial scenarios. Here, two parties in the second layer who share particles $B$ and $C$ may perform local joint operations defined by a biseparable quantum channel $\mathcal{E}_{A|BC}(\cdot)$ over the bipartition $A$ and $\{B,C\}$.}
\label{fig1}
\end{figure}

Bravyi \textit{et al.} \cite{Bra} investigated 2D Hidden Linear Function problem in terms of constant-depth quantum circuit using special quantum gates. This is further extended to other circuits over special gates \cite{Watt}. These results intrigue new ideas to explore multipartite entanglement with shallow quantum circuits. Especially, each bipartite separable state can be generated by using one layer of quantum separable channel, as Fig.\ref{fig1}(a). This suggests a novel model for generating multipartite entanglement by using different layers of biseparable completely-positive trace-preserving (BCPTP) channels \cite{Sv} from fully separable states, as shown in Fig.\ref{fig1}(b). A nature problem is what's the relationship among the entangled states, circuit depth and quantum channels.

The goal of this work is to characterize multipartite entanglement based on shallow quantum circuits of bipartite quantum channels.  We firstly prove any multipartite entanglement can be generated by using a 2-layer shallow quantum circuit consisting of two biseparable completely-positive trace-preserving channels. If the second layer consists of a convex combination of local fully separable channels with one joint channel, the present model further implies a complete hierarchy for characterizing any multipartite entanglement according to the joint ability in its generation circuits. This second layer can be further regarded as an adversarial model in cryptographic applications such as quantum secret sharing \cite{HB}, as shown in Fig.\ref{fig1}(c). The present model provides a simple method to characterize general multipartite entanglement using Schmidt numbers of reduced density matrices. Its shows a different multipartite nonlocality from previous models \cite{Luo2022,NWR,Kraft,Luo2021}.

\section*{Result}

\subsection*{Genuinely multipartite entanglement generated with shallow quantum circuits}

A general isolated $d$-dimensional quantum system is represented by a normalized vector $|\phi\rangle$ in Hilbert space $\mathcal{H}_{d}$. Instead, an open system is described by probabilistically mixing an ensemble of pure states $\{|\phi_i\rangle\}$, that is, $\rho=\sum_ip_i|\phi_i\rangle\langle\phi_i|$, where $\{p_i\}$ is a probability distribution. An $n$-particle pure state $|\Phi\rangle$ on Hilbert space $\otimes_{i=1}^n\mathcal{H}_{A_i}$ is biseparable \cite{Sv} if it can be represented by  $|\Phi\rangle=|\phi\rangle_I|\psi\rangle_{\overline{I}}$ with two pure states $|\phi\rangle$ and $|\psi\rangle$, where $I$ and $\overline{I}$ are bipartition of $\{A_1, \cdots, A_n\}$.  Note that $|\Phi\rangle$ can be generated from a fully separable state $|0\rangle^{\otimes n}$ with a 1-layer shallow circuit of biseparable quantum channel defined by $\mathcal{E}(\cdot):=U_I\otimes V_{\overline{I}}(\cdot )U_I^\dag\otimes V_{\overline{I}}^\dag$, that is,
\begin{eqnarray}
|\Phi\rangle=(U_I\otimes V_{\overline{I}})|0\rangle^{\otimes n}
\end{eqnarray}
where $U_I(\otimes_{A_i\in I}|0\rangle_{A_i})=|\phi\rangle_I$ and $V_{\overline{I}}(\otimes_{A_j\in \overline{I}}|0\rangle_{A_j})=|\psi\rangle_{\overline{I}}$. This intrigues a multipartite entanglement model in terms of shallow quantum circuits. Especially, define a biseparable completely positive trace-preserving (BCPTP) channel \cite{Werner,Sv} on Hilbert space $\mathcal{H}_I\otimes \mathcal{H}_{\overline{I}}$ as
\begin{eqnarray}
\mathcal{E}_{I}(\rho)=\sum_{i}(K_{i}\otimes S_i)\rho(K_i^\dag\otimes S_i^\dag)
\label{eq0}
\end{eqnarray}
where $K_i,S_i$ are respective Kraus operators on Hilbert space $\mathcal{H}_I$ and $\mathcal{H}_{\overline{I}}$ and satisfy $\sum_{i}K_i^\dag K_{i}\otimes S_i^\dag S_i=\openone$ with the identity operator $\openone$, $\mathcal{H}_I=\otimes_{A_i\in I}\mathcal{H}_{A_i}$, and $\mathcal{H}_{\overline{I}}=\otimes_{A_j\in \overline{I}}\mathcal{H}_{A_j}$. For a given biseparable state $\rho_{bs}=\sum_ip_i\rho_{i|I}\otimes \varrho_{i|\overline{I}}$ over a given bipartition $I$ and $\overline{I}$, it is straightforward to show from Eq.(\ref{eq0}) that there is a BCPTP channel $\mathcal{E}_{I}$ with $K_i: |0\rangle\mapsto \rho_{i|I}$ and $S_i: |0\rangle\mapsto \varrho_{i|\overline{I}}$ such that
\begin{eqnarray}
\rho_{bs}=\mathcal{E}_{I}(|0\rangle\langle0|^{\otimes n})
\end{eqnarray}
This means that any biseparable state \cite{Sv} can be generated by a probabilistically  convex combination of BCPTP channels, that is,
\begin{eqnarray}
\rho_{bs}=\mathcal{E}(|0\rangle\langle0|^{\otimes n})=\sum_{I}q_I\mathcal{E}_I(|0\rangle\langle0|^{\otimes n})
\label{eq00}
\end{eqnarray}
where the summation is over any proper subset of $\{A_1, \cdots, A_n\}$. Thus BCPTP channel provides an equivalent representation of the biseparable model \cite{Sv}. Our goal in what follows is to explore the multipartite entanglement in terms of its shallow generation circuits consisting of BCPTP channels.

Note that one layer of BCPTP channel can only build a biseparable state \cite{Sv}. This implies that the nontrivial example should be at least two layers. Especially, for any genuinely $n$-partite entanglement on finite-dimensional Hilbert spaces $\otimes_{i=1}^n\mathcal{H}_{A_i}$ \cite{Sv}, assume that the Schmidt decomposition with respect to the bipartition $I=\{A_1\}$ and $\overline{I}=\{A_2,\cdots, A_n\}$ is given by
\begin{eqnarray}
|\Phi\rangle=\sum_{i=1}^d\sqrt{\lambda_i}|\phi_i\rangle_{I}|\psi_i\rangle_{\overline{I}}, \label{eq1a}
\end{eqnarray}
where $\lambda_i$'s are Schmidt coefficients satisfying $\sum_{i=1}^d\lambda_i=1$, $\{|\phi_i\rangle\}$ are orthogonal states of $A_1$, and $\{|\psi_i\rangle\}$ are orthogonal states of all the systems in $\overline{I}$. There is a unitary transformation $U$ on Hilbert space $\mathcal{H}_{\overline{I}}$ satisfying (Supplementary note 1):
\begin{eqnarray}
U: |\psi_i\rangle_{\overline{I}}\mapsto |0\rangle_{A_2}|\hat{\psi}_i\rangle_{J}, i=1, \cdots, d
\label{eq2}
\end{eqnarray}
where $\{|\hat{\psi}_i\rangle\}$ are orthogonal states of all the systems in $J:=\{A_{3}, \cdots, A_n\}$. Thus the state $|\Phi\rangle$ is generated by a 2-layer shallow quantum circuit consisting of two BCPTP channels. By considering the probabilistic mixture of BCPTP channels, it implies a general result for multipartite entanglement on finite-dimensional Hilbert spaces.

\textbf{Theorem 1}. For $n$-partite state $\rho$ on finite-dimensional Hilbert spaces, it can be generated by a 2-layer quantum circuit given by
\begin{eqnarray}
\rho&=&\mathcal{E}_2\circ \mathcal{E}_1(|0\rangle\langle 0|^{\otimes n})
\\
&=&\sum_{I,J}p_Iq_J\mathcal{E}_J\circ\mathcal{E}_I(|0\rangle\langle0|^{\otimes n})
\label{eq2a}
\end{eqnarray}
where $\mathcal{E}_1$ and $\mathcal{E}_2$ are respective probabilistic mixtures of BCPTP channels $\mathcal{E}_I$ and $\mathcal{E}_J$, and $\{p_I\}$ and $\{q_J\}$ are probability distributions.

\begin{figure}[ht]
\begin{center}
\resizebox{225pt}{90pt}{\includegraphics{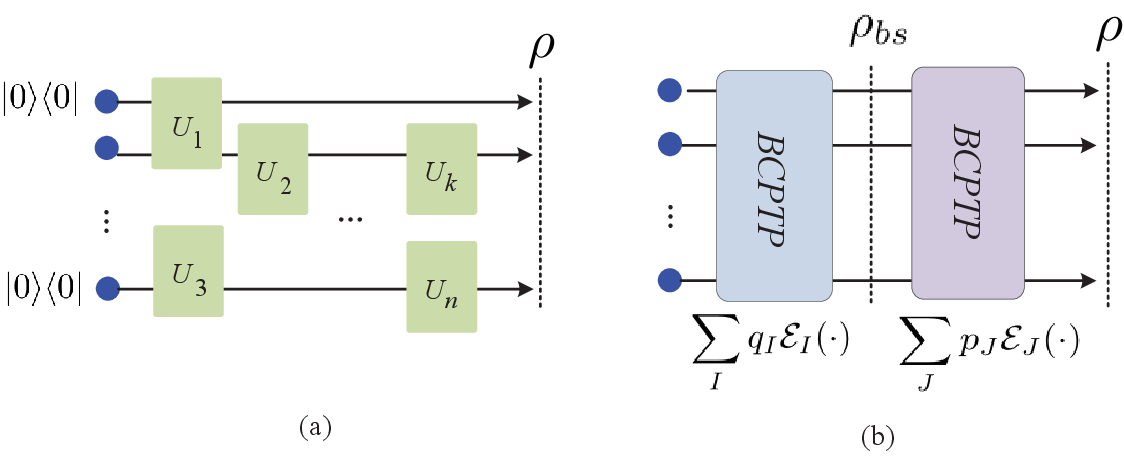}}
\end{center}
\caption{Schematic quantum circuit for generating a general state. (a) Standard quantum circuit model with different depths. Here, $U_i$ are two-particle universal logical gates \cite{Deut,Vlas}. (b) The present circuit model with two-depth. There are two BCPTP  channels $\sum_{I}p_I\mathcal{E}_I(\cdot)$ and $\sum_{J}q_J\mathcal{E}_J(\cdot)$ over different bipartitions $I,\overline{I}$ and $J,\overline{J}$.}
\label{fig2}
\end{figure}

Theorem 1 holds for any pure or mixed state on finite-dimensional Hilbert space. Eq.(\ref{eq2a}) implies a universal circuit with two depths for building any state from a fully product state, as shown in Fig.\ref{fig2}. This means any entanglement can be generated by a 2-layer shallow circuit consisting of two BCPTP channels. Note that the 2-layer quantum circuit is the smallest nontrivial shallow circuits. Theorem 1 implies the strong generation ability of small shallow circuits. This arises an immediate problem whether or not similar result holds in infinite-dimensional Hilbert spaces.

The present shallow circuit consisting of BCPTP channels is stronger than the standard quantum circuit model with an unfixed circuit depth using two-particle joint operations \cite{Deut,Vlas}. The second layer circuit of Theorem 1 means that the BCPTP channel may activate multipartite entanglement from biseparable states. This kind of entanglement swapping is the core of quantum networks \cite{BBC,Kimble}.

\subsection*{$k$-connection genuinely entanglement generated with 2-layer shallow quantum circuits}

In Theorem 1, a two-layer circuit model may be too strong both in theory and applications. The main reason is that both biseparable quantum channels allow any bipartition decomposition. Instead, we consider a one-side biseparable channel in the second layer while the first layer is to prepare a biseparable state. Here, one local joint operation $\mathcal{E}_I$ may be performed on a local set $I\subset \{A_1, \cdots, A_n\}$ while separable operations are performed on each particle in the complement set $\overline{I}$. The joint channel $\mathcal{E}_I$ can be regarded as semi-device-independent scenarios in secure applications such as quantum secret sharing, as shown in Fig.\ref{fig1}(c), where partial adversaries who own systems in $I$ may cooperate to recover other parties' information by performing joint operations while legal others are remote distributed and then not allowed to perform joint operations. Denote $\ell_I$ as the number of particles in $I$. Define a $k$-connection BCPTP channel $\mathcal{E}_I^{(k)}$ on Hilbert space $\mathcal{H}_I\otimes \mathcal{H}_{\overline{I}}$ with $\ell_I\leq k$ as
\begin{eqnarray}
\mathcal{E}_I^{(k)}(\rho)=\sum_{i}(K_{i}\otimes (\otimes_{j\in \overline{I}}S_{i;j}))\rho(K_i^\dag\otimes (\otimes_{j\in \overline{I}}S_{i;j}^\dag))
\label{eq3}
\end{eqnarray}
where $K_i$ and $S_{i;j}$ are respective Kraus operators on Hilbert space $\mathcal{H}_{I}$ and $\mathcal{H}_{A_j}$, and satisfy $\sum_{i}(K_i^\dag K_{i}\otimes (\otimes_{j\in \overline{I}}S_{i;j}^\dag S_{i;j}))=\openone$. The present $k$-connection BCPTP channel is of state-dependent. Our goal here is to explore genuinely multipartite entanglement in terms of its generation circuits with the quantum channel (\ref{eq3}) in the second layer.

\textbf{Definition 1}. An $n$-partite state is $k$-connection genuinely entanglement ($k$-CGE) if it is not a $k$-connection biseparable state given by
\begin{eqnarray}
\rho_{bs}^{(k)}=\sum_{\ell_I\leq k}p_I\mathcal{E}_I^{(\ell_I)}\circ\mathcal{E}_1(|0\rangle\langle 0|^{\otimes n})
\label{eq4}
\end{eqnarray}
where $\mathcal{E}_I^{(\ell_I)}$ are $\ell_{I}$-connection BCPTP channels in terms of the bipartition $I$ and $\overline{I}$, $\{p_I\}$ is a probability distribution, and $\mathcal{E}_1$ is a convex combination of BCPTP channels.

\begin{figure}[ht]
\begin{center}
\resizebox{200pt}{50pt}{\includegraphics{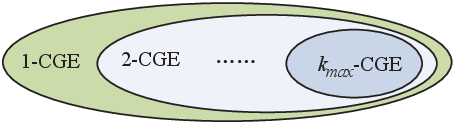}}
\end{center}
\caption{Schematic hierarchy of genuinely $n$-partite entanglement. The set consisting of $k$-CGE is included in the set consisting of $k-1$-CGE. The largest set consists of 1-CGE while the smallest set contains $k_{\max}$-CGE with $k_{\max}=\lfloor\frac{n}{2}\rfloor$, where $\lfloor x \rfloor$ denotes the maximal integer no more than $x$.}
\label{fig3}
\end{figure}

Similar to the proof of Theorem 1, the Schmidt decomposition of a given $n$-particle pure state $|\Phi\rangle$ on $d^n$-dimensional Hilbert space $\otimes_{i=1}^n\mathcal{H}_{A_i}$ is given by
\begin{eqnarray}
|\Phi\rangle=\sum_{i=1}^N\sqrt{\lambda_i}|\phi_i\rangle_{I}|\psi_i\rangle_{\overline{I}}
\label{eq5}
\end{eqnarray}
where $\lambda_i$'s are Schmidt coefficients satisfying $\sum_{i=1}^N\lambda_i=1$, $\{|\phi_i\rangle\}$ are orthogonal states of all the systems in $I$, $\{|\psi_i\rangle\}$ are orthogonal states of all the systems in $\overline{I}$. From Eq.(\ref{eq5}) we have $N\leq \min\{d^{\ell_I}, d^{n-\ell_I}\}$. This implies that $N\leq d^{n-\ell_I}\leq d^{n/2-1}$ for any integer $\ell_I$ with $\ell_I\geq n/2 +1$. There is a unitary transformation $U$ on Hilbert space $\mathcal{H}_{I}$ satisfying
\begin{eqnarray}
U: |\psi_i\rangle_{I}\mapsto |0\rangle_{A_j}|\hat{\psi}_i\rangle_{J}, i=1, \cdots, N
\label{eq6}
\end{eqnarray}
where $A_j\in \overline{I}$ and $\{|\hat{\psi}_i\rangle\}$ are orthogonal states of the particles in $J:=\overline{I}-\{A_j\}$. So, the state $|\Phi\rangle$ is an $\ell_I$-connection biseparable. This implies a general result for generating multipartite entanglement with different connection abilities as follows.

\textbf{Theorem 2}. For an $n$-partite state $\rho$ on finite-dimensional Hilbert space, it can be generated by a $2$-layer shallow quantum circuit as
\begin{eqnarray}
\rho&=&\mathcal{E}_2^{(k)}\circ \mathcal{E}_1(|0\rangle\langle 0|^{\otimes n})
\\
&=&\sum_{I,J}p_Iq_J\mathcal{E}_J^{(k)}\circ\mathcal{E}_I(|0\rangle\langle0|^{\otimes n})
\label{eq2ab}
\end{eqnarray}
where $\mathcal{E}_1$ is a convex combination of BCPTP channels $\mathcal{E}_I$ and $\mathcal{E}_2^{(k)}$ is a convex combination of $k$-connection BCPTP channels $\mathcal{E}_J^{(k)}$ defined in Eq.(\ref{eq3}) with $k\geq n/2+1$.

Theorem 2 rules out the possibility of $k$-CGE for large integer $k$. The situation is different for small $k$. For special case of $n=2$, Definition 1 is reduced to the standard separable model of bipartite systems \cite{Schr,HHH}. For each $n\geq 3$, from Definition 1 any biseparable state \cite{Sv} is $1$-connection biseparable state. In general, the connection ability $k$ is state-dependent.

\subsection*{A complete hierarchy of genuinely multipartite entanglement}

Note that any $k$-connection biseparable state is an $s$-connection biseparable state for any $s\geq k$. This implies a complete hierarchy for all the multipartite entangled states, as shown in Fig.\ref{fig3}. The largest set contains $1$-CGEs, that is, the genuinely multipartite entanglement in the biseparable model \cite{Sv}. Instead, the smallest set consists of the strongest multipartite entanglement, that is, $k_{\max}$-CGE with $k_{\max}=\lfloor \frac{n}{2}\rfloor$. Since each subset of $k$-CGE is not empty (see examples in what follows), the new classification is strict from Theorems 1 and 2, that is, each entanglement belongs to the only set $k$-CGE while it is not in $k+1$-CGE for some $k$.

For any $n$-particle pure state $|\Phi\rangle$ on a finite-dimensional Hilbert space $\otimes_{i=1}^n\mathcal{H}_{A_i}$, from Eq.(\ref{eq5}) the orthogality of $\{|\phi_i\rangle_{I}\}$ allows for constructing a unitary transformation (\ref{eq6}) if and only if its Schmidt number satisfies $N\leq d^{k-1}$ with $\ell_I=k$. This implies a directive way to find the connection-ability $k$ for a given pure state using its Schmidt numbers of reduced density matrices (Supplementary note 2).

\textbf{Theorem 3}. An $n$-partite pure state on a $d^n$-dimensional Hilbert space $\otimes_{i=1}^n\mathcal{H}_{A_i}$ is $k$-CGE if and only if the Schmidt number of the reduced density matrix of any $k$ particles is larger than $d^{k-1}$.

From Definition 1 any genuinely multipartite entanglement in the biseparable model \cite{Sv,See} is 1-CGE. Moreover, the present $k$-CGE is stronger than the robust entanglement with the robustness-depth $k$ since the particle-losing channel \cite{Luo2022} is local CPTP channel. From Theorem 3 it generally requires to evaluate Schmidt numbers of almost all the reduced density matrices of $s$-particle with $s\leq 2^n$. This yields to a NP hard problem for general entanglement because of exponential number ($O(d^{2/n})$) of reduced states. Instead, it is easy for special states.

\subsection*{Examples}

\textbf{Example 1}. An $n$-partite Greenberger-Horne-Zeilinger (GHZ) state \cite{GHZ} is given by
\begin{eqnarray}
|GHZ\rangle=\sum_{i=1}^da_i|i\cdots{}i\rangle_{A_1\cdots{}A_n}
\label{ghz}
\end{eqnarray}
where $a_i$'s satisfy $\sum_{i=1}^da_i^2=1$. It is easy to verify the state (\ref{ghz}) is 1-CGE from its permutational symmetry.

\textbf{Example 2}. Consider an $n$-partite W-type state \cite{Dur}:
\begin{eqnarray}
|W\rangle_{A_1\cdots{}A_n}=\sum_{i=1}^da_i|1_i\rangle+a_{n+1}|1\cdots 1\rangle
\label{Wstate}
\end{eqnarray}
where $|1_i\rangle$ denotes the $i$-excitation defined by $|1_i\rangle=|0\rangle^{\otimes i-1}|1\rangle|0\rangle^{\otimes n-i}$. From Theorem 3 it is easy to prove $|W\rangle$ is a 2-CGE for $a_i\not=0$, $i=1, \cdots, n+1$. This can be extended for general $n$-qudit Dicke states with $s$ excitations \cite{Dicke} given by
\begin{eqnarray}
|D_{s,n}\rangle=\frac{1}{\sqrt{L_s}}\sum_{i_1+\cdots{}+i_n=s}|i_1 \cdots{}i_n\rangle_{A_1\cdots{}A_n}
\label{eq8}
\end{eqnarray}
where $L_s$ is the normalization constant. It is a $k$-CGE with $k=\lfloor\log_d (s+1)\rfloor+1$ (Supplementary note 3). This is beyond previous models \cite{Sv,See} which do not distinguish GHZ state and W state. Moreover, the state $|D_{s,n}\rangle$ is equivalent to $|D_{(d-1)^n-s,n}\rangle$ under local unitary operations. This implies the strongest nonlocality of Dicke state $|D_{s,n}\rangle$ with $s=\lfloor(d-1)^n/2\rfloor$.

\begin{figure}[ht]
\begin{center}
\resizebox{210pt}{250pt}{\includegraphics{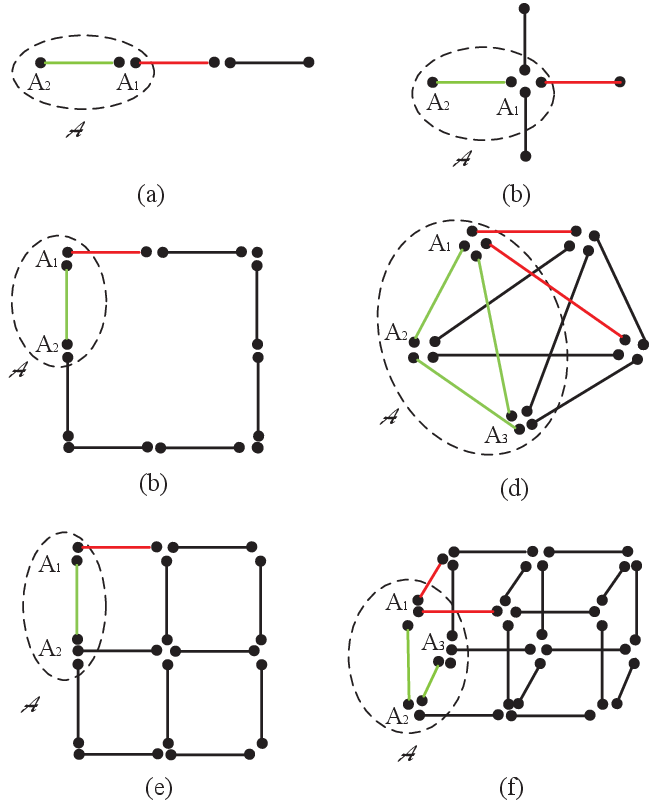}}
\end{center}
\caption{Schematic quantum networks. (a) Chain quantum network. (b) Star quantum network. (c) Cyclic quantum network. (d) Completely connected quantum network. (e) Planar quantum network (or cluster states). (d) Cubic quantum network (or cluster states). Here, each line with two dots denotes one bipartite entanglement. $\mathcal{A}$ denotes the set consisting of all parties who cooperate to disentangle the party $\textsf{A}_1$. The red lines denote the bipartite entangled states shared by $\textsf{A}_1$ with others out of $\mathcal{A}$. The green lines denote the bipartite entangled states shared by the parties in $\mathcal{A}$.}
\label{fig4}
\end{figure}

\textbf{Example 3}. Another example is entangled quantum network which may show different nonlocalities beyond single entanglement \cite{RBBB,Brun,Luo1}. As resource states of measurement-based quantum computation \cite{cluster}, the so-called cluster states \cite{cluster} may be generated by generalized Einstein-Podolsky-Rosen (EPR) states \cite{EPR}, as shown in Fig.\ref{fig4}. This kind of entanglement is 1-CGE (Supplementary note 4). Similar result can be extended for generalized graph states \cite{graph} consisting of generalized EPR \cite{EPR} and GHZ states \cite{GHZ}. Instead, some quantum networks may show different connection abilities. One example is the $n$-partite completely-connected network $\mathcal{N}_c$, where each pair shares one bipartite entanglement $|\phi_{ij}\rangle$. Recent result shows the joint state of any $k$-partite subnetwork in $\mathcal{N}_c$ is entangled for $k\geq 2$ \cite{Luo2022,Sv}. Hence, from Theorems 2 and 3 the joint state of $\mathcal{N}_c$ is a $k_{\max}$-CGE with $k_{\max}=\lfloor \frac{n}{2}\rfloor$. This means that $\mathcal{N}_c$ shows stronger nonlocality than GHZ state (\ref{ghz}) and W state (\ref{Wstate}) for any $n>4$ in the present model. This is different from the robust-entanglement model \cite{Luo2022}, where both the W state and $\mathcal{N}_c$ has the same robust-depth. Remarkably, it is converse to the recent result \cite{NWR,Kraft,Luo2021} which proves GHZ and W states have stronger entanglement beyond quantum networks. This shows a surprising feature of the genuinely multipartite nonlocality beyond bipartite scenarios, that is, it is of model-dependent.

For general quantum networks, it is general difficult to find the largest $k$ such that the total state of $\mathcal{N}_q$ is $k$-CGE. Here, we provide a polynomial-time algorithm (Algorithm 1) for estimating the upper bound of $k$ for featuring its connection ability. This is inspired by Lemmas 1 and 2 (Supplementary note 5).

\begin{algorithm}[htbp]
\caption{Verifying any $n$-partite quantum network $\mathcal{N}_q$ consisting of bipartite entanglement}
\KwIn{Finite-size network $\mathcal{N}_q$}
\KwOut{$k$, satisfying that $\mathcal{N}_q$ is at most $k$-connection entanglement}
\begin{itemize}
\item[(i)]Find the connectedness degree $\ell_i$ for any each party $\textsf{A}_i$ with $i=1, \cdots, n$.
\item[(ii)] Rearrange all parties with decreasing order into $\textsf{A}_1, \cdots, \textsf{A}_n$ (for simplicity).

\item[(iii)]Find $J$ such that $\ell_j=\min\{\ell_i\}$ with $j\in J$. Let $m=|J|$.

\item[(iv)] For $s=1: m$

\item[(v)] \quad  For $t=1: \lfloor\frac{\ell_j+1}{2}\rfloor$
\item[]  \quad (a) Let $\mathcal{A}_1=\{\textsf{A}_j\}$ and $\overline{\mathcal{A}}_t=\{\textsf{A}_1,\cdots, \textsf{A}_n\}-\mathcal{A}_t$.
\item[] \quad  (b)  Let $\mathcal{A}_{t+1}=\mathcal{A}_t\cup \{\textsf{A}_u\}$, where $\textsf{A}_u$ has shared the most bipartite entangled states with parties in $\mathcal{A}_t$ compared with other parties in $\overline{\mathcal{A}}_t$, and $\textsf{A}_u\in \overline{\mathcal{A}}_t$.

\item[]\quad (c) Evaluate $s_{j;in}$ and $t_j$.

\item[] \quad (d) If $s_{j;in}+2t_j<s_{j;out}$
\item[]  \qquad  \qquad $t\to t+1$
\item[] \qquad \quad  Otherwise
\item[]  \qquad  \qquad  Return $v_s=\min\{\ell_j,t\}$
\item[] $s\to s+1$
\item[(vi)] Output $k\leq \min\{v_1, \cdots, v_m\}$
\end{itemize}
\end{algorithm}

For each party $\textsf{A}_j$ with the minimal connectedness degree $\ell_j$, from Lemma 1 (Supplementary note 5) if $s_{j;in}+2t_j\geq s_{j;out}$ for some $t$ there is a CPTP mapping to disentangle all the particles shared by $\textsf{A}_j$. Hence, the total state of $\mathcal{N}_q$ is $t$-connection biseparable. The time complexity of the step (i) is at most $O(n^2)$. The time complexity of the step (iii) is at most $O(n)$. For a given party $\textsf{A}_j$ with $j\in J$, the time complexity of the step (b) is at most $O(n^2)$. Hence, the total time complexity is at most $O(n^4)$.

Some examples are shown in Fig.\ref{fig4}. For the chain quantum network in Fig.\ref{fig4}(a), the party $\textsf{A}_1$ shares one bipartite entanglement (red line) with other parties out of $\mathcal{A}$, and shares one bipartite entanglement (green line) with $\textsf{A}_2$. From Lemma 1 the chain quantum network in Fig.\ref{fig4}(a) is 2-connection biseparable, where the party $\textsf{A}_1$ can be disentangled with other parties out of $\mathcal{A}$ by using joint operation of $\textsf{A}_1$ and $\textsf{A}_2$. Moreover, it is genuinely multipartite entanglement \cite{Luo2021,Sv}, that is, any local operation cannot disentangle one party. Thus the chain quantum network in Fig.\ref{fig4}(a) is 1-CGE. Similar result holds for the star quantum network in Fig.\ref{fig4}(b) and cyclic quantum network in Fig.\ref{fig4}(c). For a completely connected quantum network in Fig.\ref{fig4}(d), the party $\textsf{A}_1$ shares two bipartite entangled states (red lines) with others out of $\mathcal{A}$. There are three bipartite entangled states (green lines) shared by parties in $\mathcal{A}$. From Lemma 1, it is 3-connection biseparable while any two parties cannot jointly disentangle one party. Hence, this quantum network is 2-CGE, where the party $\textsf{A}_1$ can be disentangled with others out of $\mathcal{A}$ by using joint operation of the parties $\textsf{A}_1,\textsf{A}_2$ and $\textsf{A}_3$. In general, we can prove an $n$-partite completely connected network is $k_{\max}$-CGE with $k_{\max}=\lfloor \frac{n}{2}\rfloor$. Similarly, from Lemma 2 the planar quantum network in Fig.\ref{fig4}(e) is 1-CGE while the cubic quantum network is 2-CGE.

\subsection*{Robustness of $k$-CGE}

From Eq.(\ref{eq4}) all the $k$-connection biseparable states constitutes a convex set $\mathcal{S}_k$. This allows to verify a general entanglement near to a given $k$-CGE $|\Phi\rangle$ by using linear entanglement witness \cite{HHH96,HHH} defined by
\begin{eqnarray}
\mathtt{W}_{|\Phi\rangle}=r\openone -|\Phi\rangle\langle \Phi|
\label{eq8a}
\end{eqnarray}
where $r=\max_{\rho_{cb}\in \mathcal{S}_k}D(\rho_{cb},|\Phi\rangle\langle \Phi|)$ for any distance function $D(\cdot,\cdot)$ of two states $\rho_{cb}$ and $|\Phi\rangle$ \cite{HHH}, and $\openone$ is the identity operator. From Theorem 3 the entanglement witness for any genuinely multipartite entanglement \cite{HHH,Sv} is useful for verifying a 1-CGE. One example is the GHZ state (\ref{ghz}) with $r=\max\{a_1^2,\cdots, a_d^2\}$.

\begin{figure}[ht]
\begin{center}
\resizebox{180pt}{140pt}{\includegraphics{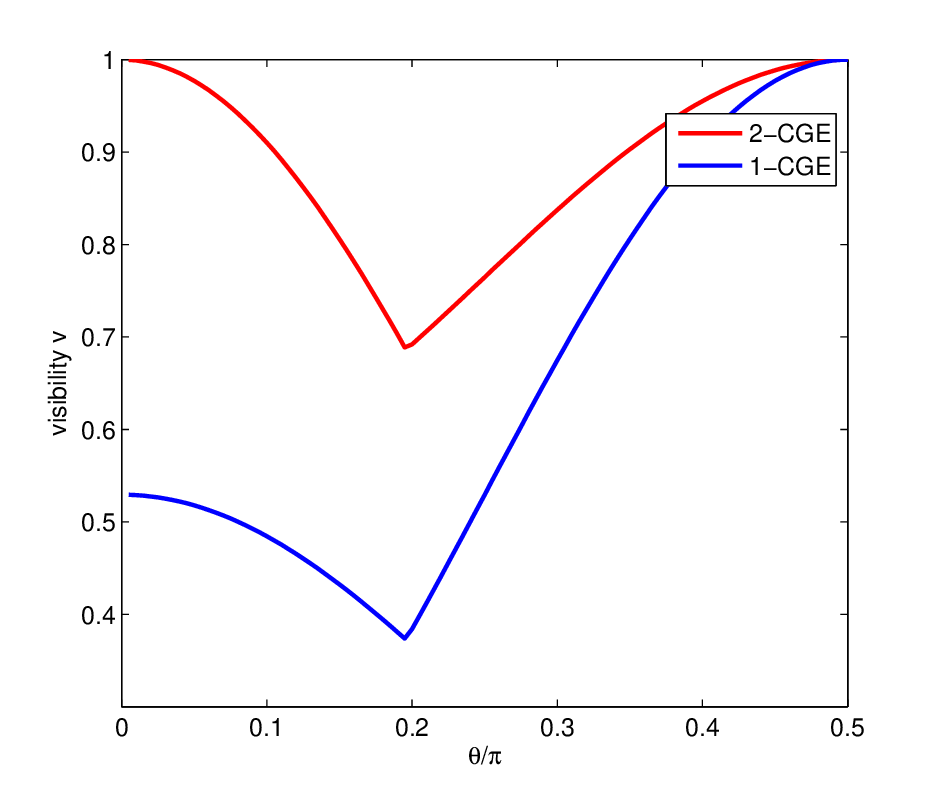}}
\end{center}
\caption{Visibilities of noisy 4-particle W state. Here, we consider a special case of $a_1=a_2=a_3=a_4=\frac{\cos\theta}{2}$ with $\theta\in (0,\frac{\pi}{2})$. In this case, the visibility for 2-CGE is given by $r=\max\{\cos^2\theta,1-\frac{\cos^2\theta}{2}\}$. The visibility for 1-CGE \cite{HHH,Sv} is given by $r_0=\max\{\sin^2\theta,\frac{\cos^2\theta}{2}\}$.}
\label{fig5}
\end{figure}

For the W-type state (\ref{Wstate}) with $n=4$, it is easy to get $r=\max\{1-a_5^2, 1-a_i^2-a_j^2, 1 \leq i<j\leq 4\}$. This implies a different entanglement witness for verifying 2-CGE beyond the genuinely multipartite entanglement (or 1-CGE) \cite{HHH,Sv} with $r=\max\{a_5^2, a_i^2+a_j^2, 1\leq i<j\leq 4\}$. Both visibilities are $v>\frac{16r+1}{17}$ for the Werner state \cite{Werner} $\rho_v=v|W\rangle\langle W|+\frac{1-v}{16}\openone$, as shown in Fig.\ref{fig5}.

\section*{Discussion}

Theorems 1 and 2 provide efficient ways for generating all multipartite entangled states with two layers of biseparable completely positive trace-preserving channels. A general problem is to explore the different channels in each layer, or different depths of shallow circuits with fixed gate sizes by using special gates \cite{Bra,Watt}. Result 3 provides an efficient way for verifying special multipartite entanglements. This intrigues a natural problem to explore new way for general multipartite entanglements. While the entanglement is the weakest nonlocality of quantum states, one may explore new hierarchies in terms of the so-called multipartite steering \cite{HR,JSU} or Bell nonlocality \cite{Bell,GHZ}. This is of special importance for recovering novel multipartite nonlocality beyond bipartite scenarios. Additionally, the present model shows the first example which shows converse multipartite nonlocality to recent quantum network model \cite{NWR,Kraft,Luo2021}. This intrigues a basic problem to explore the intrinsic nonlocality or the most reasonable model for multipartite systems.

We investigated general genuinely-generation multipartite entanglement with the help of shallow circuits. We proposed a two-layer shallow circuit model to characterize all the multipartite states in terms of biseparable completely positive trace-preserving channels. We further defined a multiple-layer circuit model with one-side local-connection. The one-side local joint operation has provided a general standard for characterizing the connecting ability of general multipartite entanglement. We obtained a simple hierarchy of multipartite entanglement in terms of the connection ability. The new entanglement witness is used to verify the entanglement robustness. These results should be interesting in multipartite entanglement theory, quantum communication, and quantum computation.

\section*{Data availability}

There is no data for the theoretical result.

\section*{Code availability}

There is no code in this paper.

\section*{SUPPLEMENTARY NOTE 1: Proof of Eq.(6)}

The Schmidt decomposition of any $n$-partite pure state $|\Phi\rangle$ on Hilbert space $\otimes_{i=1}^n\mathcal{H}_{A_i}$ is given by
\begin{eqnarray}
|\Phi\rangle=\sum_{i=1}^d\sqrt{\lambda_i}|\phi_i\rangle_{A_1}|\psi_i\rangle_{A_{2}\cdots{}A_n}
\label{eqA1}
\end{eqnarray}
where $\lambda_i$ are Schmidt coefficients satisfying $\sum_{i=1}^d\lambda_i=1$, $\{|\phi_i\rangle\}$ are orthogonal states of the particle $A_1$, and $\{|\psi_i\rangle\}$ are orthogonal states of the particles $A_{2}, \cdots, A_n$. For a given set of orthogonal states $\{|\psi_i\rangle, i=1, \cdots, d\}$, there exists an orthogonal basis $\{|\psi_j\rangle, j=1, \cdots, d\}$ of Hilbert space $\otimes_{i=2}^n\mathcal{H}_{A_i}$. Similarly, for a given set of orthogonal states $\{|\hat{\psi}_i\rangle\}$ of the particles $A_{3}, \cdots, A_n$, there exists an orthogonal basis $\{|\hat{\phi}_j\rangle\}$ of Hilbert space $\otimes_{i=2}^n\mathcal{H}_{A_i}$ such that $|\hat{\phi}_i=|0\rangle_{A_2}|\hat{\psi}_i\rangle_{A_{3}\cdots{}A_n}$ for $i=1, \cdots, d$. This can be obtained by extending the set $\{|0\rangle_{A_2}|\hat{\psi}_i\rangle_{A_{3}\cdots{}A_n}\}$ into an orthogonal basis on Hilbert space $\otimes_{i=2}^n\mathcal{H}_{A_i}$. With this basis, define the following mapping on Hilbert space $\otimes_{i=2}^n\mathcal{H}_{A_i}$ as
\begin{eqnarray}
U:&& |\psi_i\rangle_{A_{2}\cdots{}A_n}\mapsto |0\rangle_{A_2}|\hat{\psi}_i\rangle_{A_{3}\cdots{}A_n}, i=1, \cdots, d
\nonumber
\\
&&
|\psi_j\rangle_{A_{2}\cdots{}A_n}\mapsto |\hat{\phi}_j\rangle_{A_{2}\cdots{}A_n}, i=d+1, \cdots, d^{n-1}
\label{eqA2}
\end{eqnarray}
Note that $\{|\psi_i\rangle_{A_{2}\cdots{}A_n}\}$ and $\{|\hat{\phi}_j\rangle_{A_{2}\cdots{}A_n}\}$ are orthogonal bases of Hilbert space $\otimes_{i=2}^n\mathcal{H}_{A_i}$. Hence, $U$ is a unitary transformation on Hilbert space $\otimes_{i=2}^n\mathcal{H}_{A_i}$. This has completed the proof of Eq.(6) in the main text.

\section*{SUPPLEMENTARY NOTE 2: Proof of Theorem 3}

For a given $n$-partite pure state $|\Phi\rangle$ on Hilbert space $\otimes_{i=1}^n\mathcal{H}_{A_i}$, assume that the Schmidt decomposition is given by
\begin{eqnarray}
|\Phi\rangle=\sum_{i=1}^N\sqrt{\lambda_i}|\phi_i\rangle_{I}|\psi_i\rangle_{\overline{I}}
\label{B1}
\end{eqnarray}
for each bipartition $I$ and $\overline{I}$ of $\{A_1, \cdots, A_n\}$, where $\lambda_i$ are Schmidt coefficients satisfying $\sum_i\lambda_i=1$, $\{|\phi_i\rangle_{I}\}$ are orthogonal states of particles in $I$, and $\{|\psi_i\rangle_{\overline{I}}\}$ are orthogonal states of particles in $\overline{I}$. Here, $N$ is the Schmidt number of the reduced density matrix $\rho_{I}=\sum_i\lambda_i^2|\phi_i\rangle_{I}\langle \phi_i|$. If $N\leq d^{k-1}$ with $\ell_I=k$, from Theorem 2, there exists a unitary transformation $U$ on $k$ particles $A_{i_1}, \cdots, A_{i_k}$ such that $U|\Phi\rangle=|\hat{\Phi}\rangle|0\rangle_{A_{i_1}}$ for some $n-1$-particle state $|\hat{\Phi}\rangle$. This means that $|\Phi\rangle$ is $k$-connection biseparable state, that is, it is not $k$-CGE.

Moreover, $N>d^{k-1}$ with $\ell_I=k$, we show that $|\Phi\rangle$ is $k$-CGE. The proof is completed by contradiction. Assume that $|\Phi\rangle$ is not $k$-CGE, that is, $|\Phi\rangle$ is $k$-connection biseparable state. Hence, there exists a unitary operation $U$ on $k$ particles $A_{i_1}, \cdots, A_{i_k}$ such that $U|\Phi\rangle=|\hat{\Phi}\rangle|0\rangle_{A_{i_j}}$. Let $I=\{A_{i_1}, \cdots, A_{i_k}\}$. From Eq.(\ref{B1}), we have
\begin{eqnarray}
|\hat{\Phi}\rangle|0\rangle_{A_{i_j}}&=&\sum_{t=1}^N\sqrt{\lambda_t}U|\phi_t\rangle_{I}|\psi_t\rangle_{\overline{I}}
\nonumber
\\
&=&\sum_{t=1}^N\sqrt{\lambda_t}|\hat{\phi}_t\rangle_{I-\{A_{i_j}\}}|0\rangle_{A_{i_j}}|\psi_t\rangle_{\overline{I}}
\label{B2}
\end{eqnarray}
for some orthogonal states $\{|\hat{\phi}_t\rangle_{I-\{A_{i_j}\}}\}$ on the particles in $I-\{A_{i_j}\}$. However, the dimension of Hilbert space $\otimes_{s\in I-\{A_{i_j}\}}\mathcal{H}_{A_s}$ is $d^{k-1}$. So, there are at most $d^{k-1}$ orthogonal states. This is contradicted to the orthogonal state set $\{|\hat{\phi}_t\rangle_{I-\{A_{i_j}\}}, j=1, \cdots, N\}$ with $N>d^{k-1}$. It means that $|\Phi\rangle$ is $k$-connection biseparable state. This has completed the proof.

\section*{SUPPLEMENTARY NOTE 3: Cluster states}

In this section, we firstly prove any cluster state generated by any generalized EPR states \cite{EPR} is 1-CGE. And then, it will be extended for any graph state generated by any generalized EPR states \cite{EPR} and generalized GHZ states \cite{GHZ}.

Consider an $n$-partite cluster state $|C\rangle$ generated by generalized EPR states \cite{EPR} $|\phi_1\rangle, \cdots, |\phi_m\rangle$, where $|\phi_i\rangle=\cos\theta_i|00\rangle+\sin\theta_i|11\rangle$ with $\theta_i\in (0,\frac{\pi}{2})$, $i=1, \cdots, m$. Assume $|C\rangle$ is shared by $n$ parties $\textsf{A}_1, \cdots, \textsf{A}_n$. Each party $\textsf{A}_i$ can perform local controlled-phase operation $C_{j}(\theta_{i_j})$ on the shared two qubits, where $C_{j}(\theta_{i_j})$ is defined by
\begin{eqnarray}
C_{j}(\theta_{i_j})&=&|00\rangle\langle 00|+|01\rangle\langle 01|+|10\rangle\langle 10|
\nonumber
\\
&&+e^{i\theta_{i_j}}|11\rangle\langle 11|
\label{CC1}
\end{eqnarray}
One party may perform different $C_{j}(\theta_{i_j})$'s on each pair of shared qubits. Assume that each party $\textsf{A}_i$ shares $k_i$ EPR states with others. Here, all the $k_i$ qubits shared by the party $\textsf{A}_i$ is jointly combined into a particle on $2^{k_i}$-dimensional Hilbert space $\mathcal{H}_{A_i}$. Denote $\rho_{i}$ as the reduced density matrix of the party $\textsf{A}_i$. It is easy to obtain the Schmidt number of $\rho_{i}$ is $2^{k_i}$. Moreover, any local controlled-phase operation (\ref{CC1}) does not change its Schmidt number. This has proved that $|C\rangle$ is a 1-connection state. Moreover, it is genuinely multipartite entanglement \cite{Luo2021,Sv}. So, $|C\rangle$ is a 1-CGE.

Now, consider an $n$-partite graph state $|G\rangle$ generated by generalized EPR states \cite{EPR} $|\phi_1\rangle, \cdots, |\phi_m\rangle$ and generalized GHZ states \cite{GHZ} $|\psi_1\rangle, \cdots, |\psi_\ell\rangle$, where $|\psi_i\rangle$ are $k_i$-qubit GHZ states defined by $|\psi_i\rangle=\cos\vartheta_{i}|0\rangle^{\otimes k_i}+\sin\vartheta_i|1\rangle^{\otimes k_i}$ with $\vartheta_i\in (0, \frac{\pi}{2})$.  Assume each party $\textsf{A}_i$ shares $s_i$ EPR states and $t_i$ GHZ states with others. Each party $\textsf{A}_i$ can perform local joint controlled-phase operation $CC_{j}(\theta_{i_j})$ on the shared $u$ qubits, where $CC_{j}(\theta_{i_j})$ is defined by
\begin{eqnarray}
CC_{j}(\theta_{i_j})&=&\sum_{i_1\cdots i_u\not=1\cdots 1}|i_1\cdots i_u\rangle\langle i_1\cdots i_u|
\nonumber
\\
&&+e^{i\theta_{i_j}}(|1\rangle\langle 1|)^{\otimes u}
\label{CC2}
\end{eqnarray}
All the $s_i+t_i$ qubits shared by the party $\textsf{A}_i$ is jointly combined into a particle in $2^{s_i+t_i}$-dimensional Hilbert space $\mathcal{H}_{A_i}$. Denote $\varrho_{i}$ as the reduced density matrix of the party $\textsf{A}_i$. It is easy to obtain the Schmidt number of $\varrho_{i}$ is $2^{s_i+t_i}$. Moreover, any local operation (\ref{CC2}) does not change the Schmidt number. This has proved that $|G\rangle$ is a 1-connection state. Moreover, it is genuinely multipartite entanglement \cite{Luo2021,Sv}. Hence, $|G\rangle$ is $1$-CGE.

A further result for connection ability of general networks will be proved in supplementary note 5.

\section*{SUPPLEMENTARY NOTE 4: Connection ability of Dicke state}

In this section, we prove any $n$-qudit Dicke state (18) in main text is $k$-CGE with $k=\lfloor\log_d (s+1)\rfloor+1$. Specially, for any given bipartition $I=\{A_{j_1}, \cdots, A_{\ell_{I}}\}$ and $\overline{I}$ with $\ell_I=k+1$, the Schmidt decomposition of $|D_{s,n}\rangle$ is given by
\begin{eqnarray}
|D_{s,n}\rangle=\sum_{i=0}^{L_I}\sqrt{\gamma_i}|\phi_i\rangle_{I}|\psi_i\rangle_{\overline{I}}
\label{D1}
\end{eqnarray}
where $\{|\phi_i\rangle_{I}\}$ are orthogonal states of particles in $I$, $\{|\psi_i\rangle_{\overline{I}}\}$ are orthogonal states of particles in $\overline{I}$, $\gamma_i$ are Schmidt coefficients satisfying $\sum_i\gamma_i=1$, and $L_I$ denotes the Schmidt number. In fact, $|\phi_i\rangle_{I}$ are generalized $\ell_I$-qudit Dicke states defined by
\begin{eqnarray}
|\phi_i\rangle_{I}=\sum_{j_1+\cdots{}
+j_{\ell_I}=i}\alpha^{(i)}_{j_1\cdots{}j_{\ell_I}}|j_1\cdots{}j_{\ell_I}\rangle_{A_{j_1}\cdots{}A_{\ell_I}}
\label{D2}
\end{eqnarray}
where $\alpha^{(i)}_{j_1\cdots{}j_{\ell_I}}$ depend on proper coefficients of $\gamma_i$'s and satisfy $\sum_{j_1+\cdots{}
+j_{\ell_I}=i}(\alpha^{(i)}_{j_1\cdots{}j_{\ell_I}})^2=1$. Similarly, $|\psi_i\rangle_{\overline{I}}$ are generalized $n-\ell_I$-qudit Dicke states defined by
\begin{eqnarray}
|\psi_i\rangle_{\overline{I}}=\sum_{\sum_{A_{r_t}\in \overline{I}}r_t=s-i}\beta^{(i)}_{r_1\cdots{}r_{n-\ell_I}}|r_1\cdots{}r_{n-\ell_I}
\rangle_{A_{r_1}\cdots{}A_{r_{n-\ell_I}}}
\label{D3}
\end{eqnarray}
where $\beta^{(i)}_{r_1\cdots{}r_{n-\ell_I}}$ depends on some coefficients of $\gamma_i$'s and satisfy $\sum_{\sum_{A_{r_t}\in \overline{I}}r_t=s-i}(\beta^{(i)}_{r_1\cdots{}r_{n-\ell_I}})^2=1$.

Note that for each pair of $i,j$ with $i\not=j$, the states $|\phi_{i_1}\rangle$ and $|\phi_{i_2}\rangle$ are orthogonal because $\{|j_1\cdots{}j_{\ell_I}\rangle\}$ (with $j_1+\cdots{}+j_{\ell_I}=i_1$) and $\{|j'_1\cdots{}j'_{\ell_I}\rangle\}$ with $j'_1+\cdots{}+j'_{\ell_I}=i_2$) are orthogonal. This implies $\{|\phi_i\rangle_{I}\}$ defined in Eq.(\ref{D2}) are orthogonal. Similarly, we can prove $\{|\psi_i\rangle_{\overline{I}}\}$ defined in Eq.(\ref{D3}) are orthogonal. This has proved the Schmidt decomposition (\ref{D1}) with the orthogonal states in Eqs.(\ref{D2}) and (\ref{D3}). Moreover, it implies the Schmidt number $L_I=s+1$. From Theorem 3, it has proved the result.

\section*{SUPPLEMENTARY NOTE 5: Estimating upper bound of $k$ for entangled quantum networks}

In this section, we estimate the upper bound of $k$ for an $n$-partite entangled quantum network using a polynomial-time algorithm. Here, for simplicity we assume each pair shares at least one bipartite entanglement on $d$-dimensional Hilbert spaces $\mathcal{H}_d\otimes\mathcal{H}_d$.

We firstly present a sufficient and necessary condition for verifying an entangled quantum network. Consider an $n$-partite quantum network $\mathcal{N}_q$ consisting of any bipartite entanglement on Hilbert space $\mathcal{H}_d\otimes \mathcal{H}_d$. Denote $\mathcal{N}_q^{(\mathcal{A})}$ as the subnetwork consisting of all parties in $\mathcal{A}$. For a given subnetwork $\mathcal{N}_q^{(\mathcal{A})}$, denote $s_{i;in}$ denotes the inner connectedness degree of the party $\textsf{A}_i$, that is, the number of bipartite entangled states shared with other parties in $\mathcal{A}$. While $s_{i;out}$ denotes the outer connectedness degree of the party $\textsf{A}_i$, that is, the number of bipartite entangled states shared with other parties out of $\mathcal{A}$. Denote $t_i$ as all the other inner connectedness degrees in $\mathcal{N}_q^{(\mathcal{A})}$, that is, the number of bipartite entangled states shared by two parties in $\mathcal{A}-\{\mathcal{A}_i\}$. We firstly prove the following lemma.

\textbf{Lemma 1}. An $n$-partite quantum network $\mathcal{N}_q$ is $k$-connection biseparable if $s_{i;in}+2t_i\geq s_{i;out}$ for some $i$ and $k$-partite subnetwork.

\textbf{Proof of Lemma 1}. The proof is completed by two steps. One step is to transform all the entangled states shared by inner parties in a given subnetwork $\mathcal{N}_q^{(\mathcal{A})}$. The other is to change all the entangled states shared by one party into others in $\mathcal{A}$. For simplicity, consider a $k$-partite subnetwork $\mathcal{N}_q^{(\mathcal{A})}$ with $\mathcal{A}=\{\textsf{A}_1,\cdots, \textsf{A}_k\}$. In what follow, we only prove the result for the party $\textsf{A}_1$. Similar proof holds for other subnetworks and parties.

\begin{figure}[ht]
\begin{center}
\resizebox{230pt}{100pt}{\includegraphics{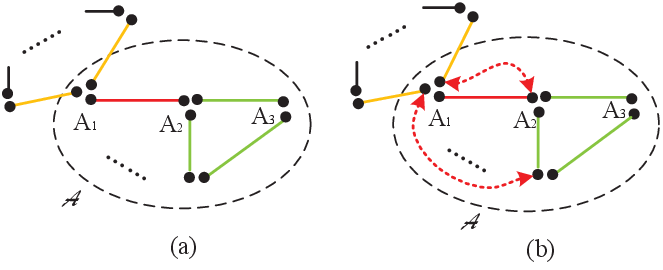}}
\end{center}
\caption{Schematic quantum network. (a) A general quantum network with subnetwork consisting of all parties in $\mathcal{A}$. Here, the orange lines denote entanglements shared by $\textsf{A}_1$ and others out of $\mathcal{A}$. The red lines denote entanglements shared by $\textsf{A}_1$ and others in $\mathcal{A}$. The green lines denote entanglements shared by parties (except for $\textsf{A}_1$) in $\mathcal{A}$. (b) Swapping two particles (red dotted lines) in $\mathcal{A}$. }
\label{figs1}
\end{figure}

Denote all the entangled states shared by the party $\textsf{A}_1$ and parties in $\mathcal{A}$ as (see red lines in Fig.\ref{figs1}(a))
\begin{eqnarray}
\varrho_{1;in}=\otimes_{i\in I}\rho_{A_{1\to i}A_{i\to 1}}
\label{E1}
\end{eqnarray}
where $\rho_{A_{1\to i}A_{i\to 1}}$ are bipartite entangled states on Hilbert space $\mathcal{H}_d\otimes \mathcal{H}_d$. Denote all the entangled states shared by the party $\textsf{A}_1$ and parties out of $\mathcal{A}$ as (see orange lines in Fig.\ref{figs1}(a))
\begin{eqnarray}
\varrho_{1;out}=\otimes_{j\in J}\rho_{A_{1\to j}A_{j\to 1}}
\label{E2}
\end{eqnarray}
where $\rho_{A_{1\to j}A_{j\to 1}}$ are bipartite entangled states on Hilbert space $\mathcal{H}_d\otimes \mathcal{H}_d$. Denote all the entangled states shared by any two  parties in $\mathcal{A}_i:=\mathcal{A}-\{\textsf{A}_1\}$ as (see green lines in Fig.S\ref{figs1}(a))
\begin{eqnarray}
\varrho_{\mathcal{A}_i}=\otimes_{A_iA_j\in \mathcal{A}_i}\rho_{A_{i\to j}A_{j\to i}}
\label{E3}
\end{eqnarray}
where $\rho_{A_{i\to j}A_{j\to i}}$ are bipartite entangled states on Hilbert space $\mathcal{H}_d\otimes \mathcal{H}_d$. There exists a local CPTP mapping $\mathcal{E}_{\mathcal{A}_i}$ such that
\begin{eqnarray}
\mathcal{E}_{\mathcal{A}_i}:\varrho_{\mathcal{A}_i}\mapsto \otimes_{A_iA_j\in \mathcal{A}_i }|00\rangle_{A_iA_j}\langle 00|
\label{E4}
\end{eqnarray}

Denote $\mathcal{A}_1$ consists of all particles $A_{1\to j}$ in Eq.(\ref{E2}). Denote $\mathcal{B}$ consists of all particles $A_{1\to j}$ in Eq.(\ref{E1}) and all particles $A_{i\to j}$ and $A_{j\to i}$ in Eq.(\ref{E2}). Let $SWAP(A_iB_i)$ be swapping operation of two particles $A_i$ and $B_i$ defined by
\begin{eqnarray}
SWAP(A_iB_i)=\sum_{i}|ii\rangle\langle ii|+\sum_{i\not=j}|ij\rangle\langle ji|
\label{Swap}
\end{eqnarray}
where $A_i\in \mathcal{A}_1$ and $B_i\in \mathcal{B}$, as shown in Fig.S\ref{figs1}(b). After these swapping operations, all the bipartite entangled states shared with $\mathcal{A}_1$ are disentangled with all the particles in $\mathcal{A}_1$, but re-entangled with particles in $\mathcal{B}$ if $s_{i;in}+2t_i\geq s_{i;out}$. This is the basic fact of entanglement swapping \cite{BBC}. Hence, if $s_{i;in}+2t_i\geq s_{i;out}$, there is a CPTP mapping $\mathcal{E}_{\mathcal{A}}=SWAP_{\mathcal{A}}\circ\mathcal{E}_{\mathcal{A}_1}$ satisfying \begin{eqnarray}
\mathcal{E}_{\mathcal{A}}:\rho_{A_1\cdots A_n}\mapsto |0\rangle_{A_1}\langle0|\otimes \hat{\rho}_{A_2\cdots{}A_n}
\label{E5}
\end{eqnarray}
where $\rho_{A_1\cdots A_n}$ denotes the total state of $\mathcal{N}_q$, $\hat{\rho}_{A_2\cdots{}A_n}$ denotes a proper state of subnetwork consisting of $\textsf{A}_2,\cdots, \textsf{A}_n$, and $SWAP_{\mathcal{A}}=\otimes_{A_i\in \mathcal{A}_1}SWAP(A_iB_i)$. This means that there is a local $k$-partite mapping for transforming $\mathcal{N}_q$ into a disconnected network. Hence, $\mathcal{N}_q$ is $k$-connection biseparable state. This has completed the proof. $\Box$

\begin{figure}[ht]
\begin{center}
\resizebox{110pt}{100pt}{\includegraphics{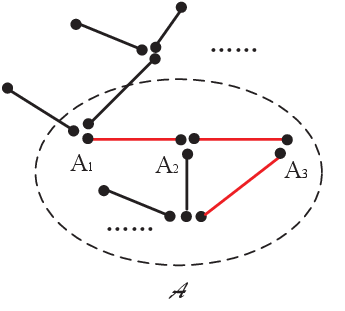}}
\end{center}
\caption{Schematic quantum network. A general quantum network with subnetwork consisting of all parties in $\mathcal{A}$. Here, there is a chain subnetwork shown in red lines. }
\label{figs2}
\end{figure}

\textbf{Lemma 2}. An $n$-partite $c$-connected quantum network $\mathcal{N}_q$ is $k$-connection biseparable state with $2k\geq c+1$, where the $c$-connectedness means that there are $c$ numbers of different chain subnetworks connecting each pair $(\textsf{A}_i,\textsf{A}_j)$ of $\mathcal{N}_q$.

In Lemma 2, a chain subnetwork consists of $\textsf{A}_{i_1}, \cdots, \textsf{A}_{i_s}$ such that each adjacent pair share one entanglement.

\textbf{Proof of Lemma 2}. For any $n$-partite $c$-connected quantum network $\mathcal{N}_q$, there is one party ($\textsf{A}_1$ for example) who shares $c$ bipartite entangled states with others. Moreover, for any $c$ and $k$ there exists a $k$-partite subnetwork $\mathcal{N}_q^{(\mathcal{A})}$ with $\textsf{A}_1\in \mathcal{A}$ such that all the parties (except for $\textsf{A}_1$) in $\mathcal{A}$ have at least one chain subnetwork, as shown red lines in Fig.S\ref{figs2}. This implies that
\begin{eqnarray}
s_{1;in}+2t_1\geq 2k-2
\label{E6}
\end{eqnarray}
where an $m$-partite chain network has $2m-2$ particles. From Lemma 1 and $s_{1;out}=c-1$ we have completed the proof. $\Box$

\section*{Acknowledgements}

We thanks for Chen Jing-Ling and Bu Kaifeng. This work was supported by the National Natural Science Foundation of China (Grants Nos. 62172341, 61772437, 12075159, 12171044), Beijing Natural Science Foundation (Grant No.Z190005), Academy for Multidisciplinary Studies, Capital Normal University, Shenzhen Institute for Quantum Science and Engineering, Southern University of Science and Technology (Grant No. SIQSE202105), and the Academician Innovation Platform of Hainan Province.

\section*{Author contributions}

M.X.L. and S.M.F. conceived the idea. M.X.L. wrote the majority of the paper and S.M.F. reviewed this main results.

\section*{Ethics declarations}

\subsection*{Competing interests}

The authors declare no competing interests.

\section*{Additional information}

Supplementary information is available for this paper online.

Correspondence and requests for materials should be addressed to M.X.

\end{document}